\documentclass[twocolumn,groupedaddress,superscriptaddress]{revtex4}
\usepackage{graphicx}
\usepackage[title]{appendix}

\usepackage{bm}
\usepackage{color}

\begin{document}

\title{Negative optical torque on a microsphere in optical tweezers}

\author{K. Diniz}
\affiliation{Instituto de F{\'i}sica, Universidade Federal do Rio de Janeiro, CP 68528, Rio de Janeiro RJ 21941-909, Brazil}
\affiliation{LPO-COPEA, Instituto de Ci\^encias Biom\'edicas, Universidade Federal do Rio de Janeiro, Rio de Janeiro, RJ, 21941-590, Brasil}
\author{ R. S. Dutra}
\affiliation{LISComp-IFRJ, Instituto Federal de Educa\c{c}\~ao, Ci\^encia e Tecnologia, Rua Sebasti\~ao de Lacerda, Paracambi, RJ, 26600-000, Brasil}
\author{ L. B. Pires}
\affiliation{Instituto de F{\'i}sica, Universidade Federal do Rio de Janeiro, CP 68528, Rio de Janeiro RJ 21941-909, Brazil}
\affiliation{LPO-COPEA, Instituto de Ci\^encias Biom\'edicas, Universidade Federal do Rio de Janeiro, Rio de Janeiro, RJ, 21941-590, Brasil}
\author{ N. B. Viana}
\affiliation{Instituto de F{\'i}sica, Universidade Federal do Rio de Janeiro, CP 68528, Rio de Janeiro RJ 21941-909, Brazil}
\affiliation{LPO-COPEA, Instituto de Ci\^encias Biom\'edicas, Universidade Federal do Rio de Janeiro, Rio de Janeiro, RJ, 21941-590, Brasil}
\affiliation{CENABIO - Centro Nacional de Biologia Estrutural e Bioimagem, Universidade Federal do Rio de Janeiro, Rio de Janeiro, RJ, 21941-902, Brasil.}
 \author{H. M. Nussenzveig}
\affiliation{Instituto de F{\'i}sica, Universidade Federal do Rio de Janeiro, CP 68528, Rio de Janeiro RJ 21941-909, Brazil}
\affiliation{LPO-COPEA, Instituto de Ci\^encias Biom\'edicas, Universidade Federal do Rio de Janeiro, Rio de Janeiro, RJ, 21941-590, Brasil}
\affiliation{CENABIO - Centro Nacional de Biologia Estrutural e Bioimagem, Universidade Federal do Rio de Janeiro, Rio de Janeiro, RJ, 21941-902, Brasil.}
\author{P. A. Maia Neto}
\affiliation{Instituto de F{\'i}sica, Universidade Federal do Rio de Janeiro, CP 68528, Rio de Janeiro RJ 21941-909, Brazil}
\affiliation{LPO-COPEA, Instituto de Ci\^encias Biom\'edicas, Universidade Federal do Rio de Janeiro, Rio de Janeiro, RJ, 21941-590, Brasil}
\affiliation{CENABIO - Centro Nacional de Biologia Estrutural e Bioimagem, Universidade Federal do Rio de Janeiro, Rio de Janeiro, RJ, 21941-902, Brasil.}

\begin{abstract}
We show that the optical
force field in optical tweezers with elliptically polarized beams
has the opposite handedness for a wide range of particle sizes and for the most common configurations. 
 Our method
is based on the direct observation of the particle equilibrium position under the effect of a
transverse
 Stokes drag force, and its rotation around the optical axis by the mechanical effect of the optical torque. We find overall agreement with theory, with no fitting, provided that astigmatism, which is characterized separately, is included in the theoretical description. Our work opens the way for characterization of the trapping
parameters, such as the microsphere complex refractive index and the astigmatism of the optical system, from measurements of the microsphere rotation angle.
\end{abstract}

%%%%%%%%%%%%%%%%%%%%%%%%%%  body  %%%%%%%%%%%%%%%%%%%%%%%%%%

\maketitle

\section{Introduction}
Negative optical forces are at the origin of single-beam optical traps \cite{Ashkin86}, also known as optical tweezers, which have 
become extremely important tools in several fields of physics \cite{Phys}  and cell biology \cite{Block2011}. 
The optical torque (OT) on a transparent and isotropic microsphere at its equilibrium position vanishes~\cite{Marston84,Barton89}, 
since Mie scattering conserves 
optical angular momentum (AM)  when the microsphere is aligned along the beam symmetry axis~\cite{Schwartz2006}.
However, if the microsphere is displaced laterally, 
optical AM might be  transferred to the microsphere center of mass \cite{Grier2012}. 

In this paper, we show, both theoretically and experimentally, that 
the resulting orbital optical torque (OT) 
points along the direction opposite to the incident field AM in most situations involving practical applications of optical tweezers with
 circularly or elliptically polarized Gaussian trapping beams.

Negative or left-handed OT is  analogous  to the negative optical force in single-beam traps. 
An oblate spheroid was predicted to spin around the axis of a circularly polarized Gaussian beam with the opposite handedness of the incident optical AM 
\cite{Simpson2007}.
Additional proposals for implementation include the employment of chiral media \cite{Canaguier2015, Li2017} and particle arrays with 
discrete rotational symmetry \cite{Chen2004}.
A negative OT was theoretically predicted for a small isotropic particle illuminated by a
 vortex beam in the Rayleigh scattering approximation \cite{nieto-vesperinas2015}.
Reverse orbiting of non-spherical particles confined on a 2D interface was observed by employing  Laguerre-Gaussian vortex beams \cite{Jesacher2006}. 
Negative OT was demonstrated for a macroscopic inhomogenous and anisotropic  disk 
 by measuring the rotationally Doppler-shifted
 reflected light  \cite{Hakobyan14,Hakobyan15}
and more recently by the direct observation of the disk rotation \cite{Magallanes18}.

A pioneering paper  \cite{Friese1998} demonstrated 
positive OT on optically trapped birefringent particles
by elliptically polarized light. 
An extension to  isotropic transparent nonspherical particles was reported in~\cite{Bishop2003}.
Important applications to cell biology are reviewed in \cite{Forth2013}.
 Conversion from optical spin AM to
  mechanical orbital AM
 has been demonstrated
by focusing a Gaussian beam with circular  polarization~ \cite{Zhao2009}. 
 Positive or negative OT was observed depending on
the configuration of an array of particles  \cite{Adachi2007, Sule2017, Han2018}.
The optical force field is  non-conservative on account of 
the azimuthal force component responsible for the OT  and the resulting  Brownian motion is a rich platform for  investigating  non-equilibrium effects \cite{Grier2012,Svak2018}.

Here we measure the mechanical effect of the OT by monitoring the 
 microsphere equilibrium position 
laterally displaced by an external Stokes drag force, as a function of the helicity of the elliptically polarized trapping beam. 
Since the particle is displaced off-axis, an OT appears, and as a consequence the equilibrium position is rotated around the beam axis. 
Our experimental and theoretical results indicate that the OT turns out to be negative in most cases of practical interest provided that the trapping beam is not very astigmatic.

\section{Experimental procedure}

\begin{figure}
	\begin{center}
		\includegraphics[width=10cm]{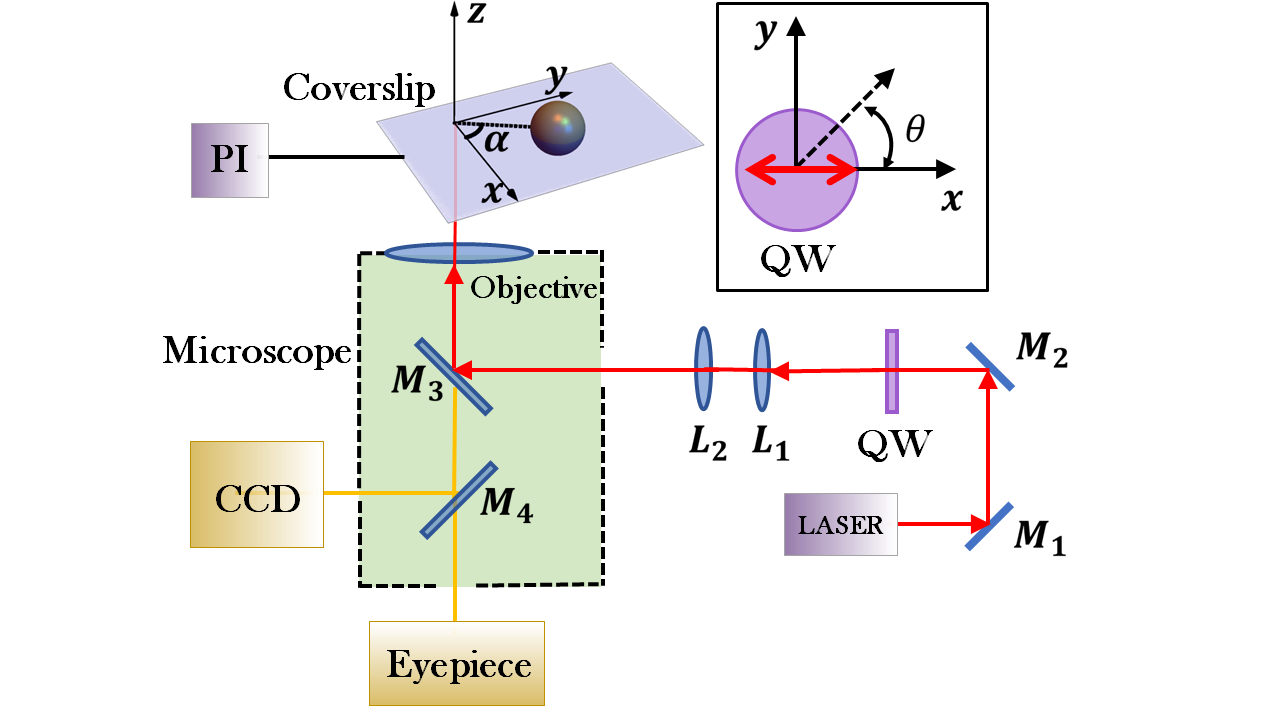}\end{center}
	\caption{ Schematic diagram of the experimental setup. The  sample is displaced along the $x$ direction. Because of the OT, the microsphere equilibrium position
	is displaced off-axis along a direction rotated by an angle $\alpha.$
	The off-axis displacement, which is smaller than the radius, is exaggerated for clarity. 
	The inset shows the angle $\theta$ between the fast axis of the quarter-wave plate (QW), represented by a dashed line, 
	and the  polarization direction of the laser beam along the $x$ axis.}
   \label{setup}
\end{figure}

Fig.~\ref{setup} is a diagram of our experimental setup.
A  ${\rm TEM}_{00}$ laser beam (IPG photonics, model YLR-5-1064LP) with wavelength $\lambda_0 = 1064\, {\rm nm}$, linearly polarized along
 the $x$ direction,  goes  through a quarter waveplate (QW) after reflection by mirrors $M_1$ and $M_2,$
that are employed
  to control the beam position.
  The fast axis of the QW makes an angle $\theta$ with the $x$ direction and the beam propagates along the positive $z$ direction.
   The laser beam is then directed to a beam expander, consisting of lens $L_1$ with focal distance $f_1 = (19.0 \pm 0.1)\, {\rm mm}$ and lens $L_2$ with focal distance $f_2 = (50.2\pm 0.1)\, {\rm mm}$. 
    The expanded beam, with waist $w = (5.82 \pm 0.08)\, {\rm mm}$ is reflected by a dichroic mirror ($M_3$) and directed to the entrance of a
    Nikon PLAN APO, 100x, NA = 1.4,
     oil-immersion objective with back aperture radius $R_{\rm obj}=(3.15 \pm 0.05)\, {\rm mm}$, of a Nikon TI-U  infinity-corrected inverted microscope. 
The total objective transmittance \cite{Viana06}, including the apodization loss due to overfilling the objective back aperture, is $(7,3 \pm 0,6)\%.$  
    The beam is focused onto a sample chamber topped by an O-ring, placed on a coverslip, containing a water suspension
     of polystyrene microspheres. 
Control of the  microsphere height is critical for 
      comparison with theory because of the spherical aberration introduced by refraction at the glass-water interface~\cite{Torok95}.
      We follow the procedure 
     of~\cite{NathanAPL,Viana2007}.
   We first move the objective down until the trapped microsphere touches the coverslip. 
     Starting from this configuration, 
     we then displace the objective upwards by distances $(3.0 \pm 0.5)\,\mu {\rm m}$ and $(7.0 \pm 0.5)\,\mu {\rm m}$
   when trapping microspheres of radii  $a=(0.50 \pm 0.02)\,\mu {\rm m}$  and $a=(1.50 \pm 0.04)\,\mu {\rm m},$ respectively. 
 The laser beam power at the objective entrance port was  $(352 \pm 1)\,{\rm mW}$ for the smaller sphere and   $(1140 \pm 3)\,{\rm mW}$ for the larger one.

\begin{figure}
	\begin{center}
		\includegraphics[width=8.cm]{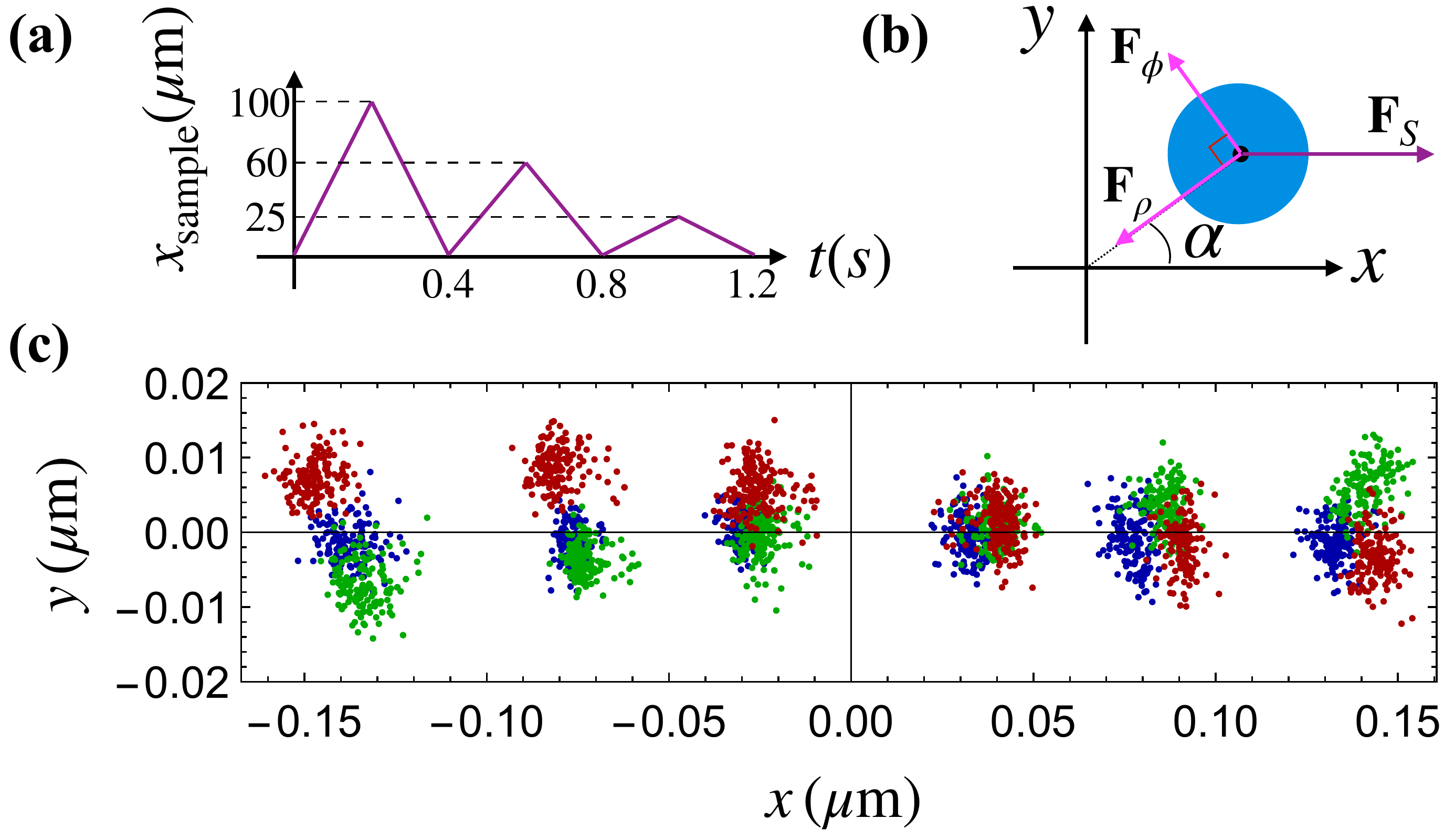}\end{center}
	\caption{
	(a) The sample is driven by the  triangle wave $x_{\rm sample}(t)$ that is repeated over time. 
	(b) The resulting Stokes drag force ${\bf F}_S$ displaces the microsphere off-axis, allowing for the transfer of optical AM.
	The equilibrium position is rotated by an angle $\alpha =\arctan(F_\phi/F_\rho),$ with  $F_{\phi}$ and $F_{\rho}$ denoting  cylindrical optical force components.
	(c) Microsphere positions on the $xy$ plane under the effect of a Stokes drag force ${\bf F}_S$ along the $x$ direction. Each lump of points corresponds to a given
	value for  ${\bf F}_S$ defined by the
	 sample velocity.
	Data are shown for
	three different values of the QW angle $\theta:$
	  $0^{\rm o}$ (blue),  $45^{\rm o}$ (green)  and $135^{\rm o}$ 
	(red), corresponding to linear, circular $\sigma^-$ and 
 circular $\sigma^+$ polarizations, respectively. The microsphere radius is $a = (0.50 \pm 0.02)\,\mu{\rm m}$.
	 The rotation angle $\alpha$ 
	is obtained from a linear fit of all points corresponding to a given $\theta.$ }
   \label{xy}
\end{figure}

In order to displace the microsphere off-axis and produce an OT, 
     the sample is alternately driven
     along  positive and negative $x$ directions
      by a piezoelectric nano-positioning stage (Digital Piezo Controller E-710, Physik Instrumente). Three different speeds are 
     employed: $ 500\, \mu {\rm m /s}, 300\, \mu {\rm m /s}$ and
        $125\, \mu {\rm m /s}, $
        defining the cycle shown in  Fig.~\ref{xy}(a) that is repeated over time. 
     Images of the entire process are recorded by a CMOS camera (Hamamatsu Orca-Flash2.8 C11440-10C) for data analysis.

     The resulting Stokes drag force
     displaces  the trapped microsphere off-axis. Since the trapping beam transfers AM to the sphere, 
     its equilibrium position 
   gets displaced to
       a direction on the $xy$ plane rotated around the $z$ axis by 
 an angle $\alpha$ with respect to the $x$ axis, as depicted in Fig.~\ref{xy}(b).
In Fig.~\ref{xy}(c), we show experimental data 
 for the  microsphere position on the $xy$ plane,
 for a microsphere of radius $a=(0.50 \pm 0.02)\,\mu {\rm m}$ and 
$\theta=0$ (blue),  $45^{\rm o}$ (green)  and $135^{\rm o}$ (red).
Points corresponding to a given sample speed appear to be lumped together as expected. 
We measure $\alpha$ from a linear fit of all points corresponding to a given $\theta$  (the transverse stiffness can also be obtained and the results are compatible with theoretical MDSA+ values \cite{Dutra2012,Dutra2014}).
 We intentionally rotate the camera by $\sim 30^{\rm o}$ with respect to the axes of the microscope stage so as to have comparable displacements along  $x$ and $y$ axes. We then 
determine an offset for $\alpha,$ representing the angle of rotation of the camera with respect to the
 direction of the sample displacement,
  by taking the 
average of the values found for $\theta=0^{\rm o},$  $90^{\rm o}$ and $180^{\rm o}$ (linear polarization), which are measured twice as often as for the other 
values of $\theta$ corresponding to non-zero helicities.
 The axes shown in Fig.~\ref{xy}(c) were rotated accordingly.

 The optical force 
    balancing the 
    Stokes force has radial and azimuthal cylindrical components 
     $F_\rho$ and $F_\phi,$ with $F_\phi$ accounting for the OT,  as illustrated in Fig.~\ref{xy}(b).
     The rotation angle is given by
     \(
 \alpha =\arctan(F_\phi/F_\rho).
\) 
We compare the experimental results for the rotation angle $\alpha$  with  theoretical results for the optical force components.
 The ratio between
 the cylindrical optical force components is calculated
with the help of the Mie-Debye  spherical aberration (MDSA) theory of optical tweezers \cite{Viana2007}, 
either with (MDSA+)~\cite{Dutra2012,Dutra2014}
 or without the inclusion of astigmatism, 
 as discussed in the next  section.

\section{MDSA theory}

 The Mie-Debye theory of optical tweezers \cite{MaiaNeto00,Mazolli03} relies on the  electromagnetic generalization of Debye's scalar model for a diffraction-limited focused optical beam derived by 
Richards and Wolf~\cite{RichardsWolf59}. 
The focused laser beam incident on the microsphere is decomposed into an angular spectrum of plane waves and the results are valid well beyond the validity range of the 
 paraxial approximation.
 The standard Mie scattering results are adapted for a general direction of incidence 
 by developing the multipolar expansion of a plane wave propagating along an arbitrary direction 
 with the help of the Wigner rotation matrix elements. The  scattered field is then obtained as a superposition of the 
contributions from all plane-wave components of the non-paraxial focused beam. 
The optical force on the sphere
${\bf F}$ 
 is finally obtained  by integrating the Maxwell stress tensor over a spherical surface at infinity.

%modif: note inversao de paragrafo 
Within MDSA, 
the model for the nonparaxial focused beam is built from the expression for the paraxial beam at the objective entrance port, 
taking into account the spherical aberration  introduced by refraction at the planar interface between the glass slide and sample aqueous solution.
We assume the laser beam at the entrance port to be
Gaussian with a planar
  wavefront.
 The elliptical polarization at the entrance port is defined by the angle $\theta$ between the fast axis of QW and the original linear polarization direction along the $x$ axis, 
 as depicted in the inset of Fig.~1:
\begin{eqnarray} \label{e1}
{\bf E}_{\rm port}(\rho,\phi,z) = E_p\, e^{i k_0 z} e^{-\frac{\rho^2}{w^2}}
\sum_{\sigma=+,-} \frac{1-ie^{-2i\sigma\theta}}{2} \;
\boldsymbol{\hat\epsilon}_{\sigma}
 \label{polarization}
\end{eqnarray}
where $E_p$ is electric field amplitude at the center of the objective entrance port and $k_0=2\pi/\lambda_0.$
$\boldsymbol{\hat\epsilon}_{\sigma}= ({\bf \hat x} +i\sigma {\bf \hat y})/\sqrt{2}$ are the unit vectors for 
right-handed ($\sigma=-1$) and left-handed ($\sigma=+1$) circular  polarizations, which are obtained by taking
the fast axis at $\theta = \pi /4$ 
and  $\theta = 3\pi /4,$
 respectively.
 
The multipole expansion of the focused beam, which plays the role of the incident field on the microsphere, is simpler for 
 circular polarization.
A general polarization state can 
be readily obtained by writing the incident field as a superposition of 
$\sigma^+$ and $\sigma^-$ circular polarizations, as illustrated by Eq.~(\ref{polarization}) for the elliptical polarization produced by QW
(see also \cite{Dutra2007} for a detailed derivation for linear polarization). 
Since the Maxwell stress tensor is quadratic in the electric and magnetic fields, ${\bf F}$ contains 
cross terms in addition to the result for  circular polarization. 
For the elliptical polarization (\ref{polarization}), the azimuthal force component 
when the sphere center is at  position $(\rho,\phi,z)$
is given by
\begin{equation}\label{Fphi}
F_{\phi} = -F_{\sigma^+}(\rho,z)\,\sin(2\theta) + F_{\rm cr}(\rho,z) \,\cos(2\theta)\,\sin[2(\phi-\theta)].
\end{equation}
 The first term on the right-hand-side of (\ref{Fphi}) accounts for 
 the transfer of  spin AM to the microsphere center-of-mass,  and
  $F_{\sigma^+}$ is the azimuthal force component when taking circular polarization $\sigma^+$ ($\theta=3\pi/4$). 
  Its explicit partial-wave expansion is given in \cite{Viana2007}.  
  The transfer of AM is modulated by 
  the optical helicity $-\sin(2\theta)$  and is 
 simply reversed when 
   replacing  $\theta\rightarrow -\theta$ as expected. 
In the particular case of circular polarization, 
the focused spot is rotationally symmetric around the $z$ axis
  and all cylindrical force components are independent of $\phi.$ 
  
   On the other hand, when the polarization at the entrance port is elliptical, 
   the non-paraxial focal spot (electric energy density map) is elongated along the major axis of the polarization ellipse (a similar effect is discussed in \cite{RichardsWolf59} for linear polarization). 
   In our case, the ellipse principal axes lie along the directions $\theta$ and $\theta+\pi/2$ defined by QW, and as a consequence 
   the gradient of the electric energy density  contains an azimuthal component depending on $\phi-\theta$ in addition to the usual radial one. 
 Thus, the gradient is radial only along the directions of the major axes: $\phi-\theta = n\pi/2$ with $n$ integer. 
   Although the optical force is not in general proportional to the gradient of the electric energy density (due to the contribution of higher multipoles in addition to the electric dipole one), the same symmetry properties of the optical force field still hold as in the Rayleigh limit. Thus, the nonparaxial spot elongation is the source of the second  term on the 
   r.-h.-s. of (\ref{Fphi}), which indeed vanishes for  $\phi-\theta = n\pi/2$  and in the case of circular polarization. 
  More specifically,  $F_{\rm cr}(\rho,z)$ results from the cross contribution $\sigma^+\cdot \sigma^-$ obtained when computing the optical force from the Maxwell stress tensor as discussed previously.
Its explicit partial-wave expansion is given in Appendix A.

\section{Results}

\begin{figure}
	\begin{center}
		\includegraphics[width=6.5cm]{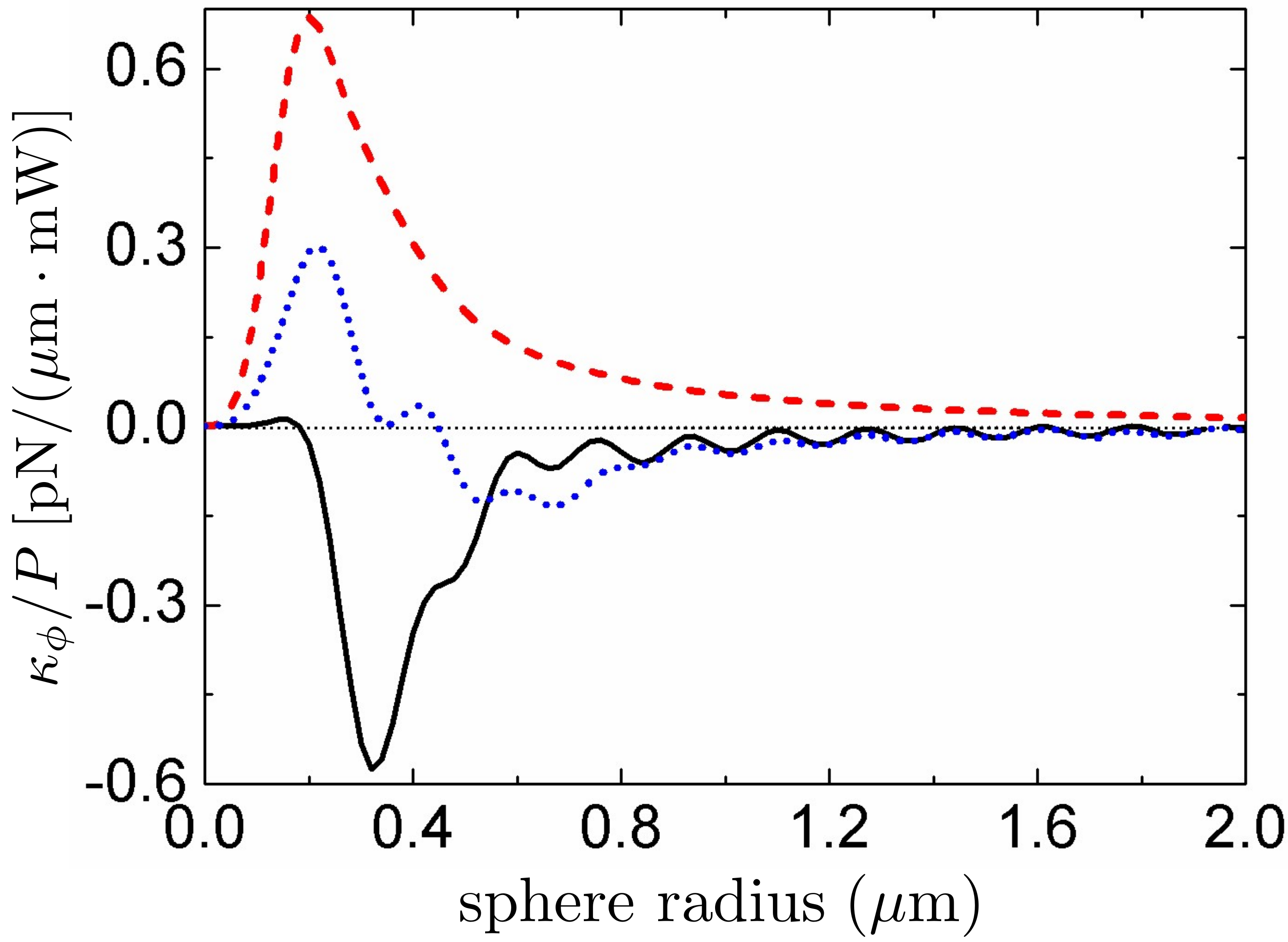}
	\end{center}
	\caption{
	Optical torsion constant $\kappa_\phi$ per unit power versus
	 microsphere radius. 
	The laser beam is left-handed ($\sigma^+$) circularly polarized.
	MDSA  (zero astigmatism) results for polystyrene (solid) and for an absorptive material 
	 with ${\rm Im}(n_{\rm sphere})= 0.5$ (dashed). 
	  MDSA+ results for polystyrene with  astigmatism amplitude $A_{\rm ast}=0.25$   (dotted).	}
 \label{plot_kappa}
\end{figure}

We start with the theoretical results derived within MDSA and MDSA+ for the parameters and conditions corresponding to our experiment. 
  As shown in Fig.~\ref{xy}(c), they correspond to  small off-axis displacements, $\rho < a,$ allowing us to
  approximate $F_{\phi}(\rho,\phi,z)\approx \kappa_\phi(\phi,z)  \rho$ in terms of
  the torsion constant, defined as
\(
\kappa_\phi= (\partial F_\phi/ \partial \rho) |_{\rho=0}.
\)
Note that the orbital OT vanishes when the sphere center is aligned along the beam symmetry axis ($\rho=0$). On the other hand, even in this case
AM may still be transferred and lead to spinning of absorbing spheres, which
is outside our current experimental reach. Therefore, it is not
considered in this paper.
In Fig.~\ref{plot_kappa}, we plot 
the variation of $\kappa_\phi$ (in units of the laser beam power $P$
in the sample region)
with the microsphere radius $a$, at the focal plane $z=0,$ for left-handed ($\sigma^+$) circular polarization  at the objective entrance port. 

In all three examples shown in Fig.~\ref{plot_kappa}, we have taken
 ${\rm Re}(n_{\rm sphere})=1.576 $ for the real part of the microsphere refractive index,  corresponding to 
  polystyrene 
at $\lambda_0 = 1064\,{\rm nm}$
 \cite{Ma03}. The dashed and solid lines represent the results  of MDSA theory, 
 which neglects optical aberrations other than
the  spherical aberration introduced  by refraction at the glass slide planar interface.
We assume the paraxial focal plane to be at a distance $3a$ from the glass slide.

When the imaginary part of the refractive index is sufficiently large, we expect 
$\kappa_\phi$ to be positive for all microsphere radii, since the positive optical AM is then transferred to 
the microsphere by absorption. We illustrate this case by taking 
 ${\rm Im}(n_{\rm sphere})= 0.5$ 
 (dashed line). 
 On the other hand, for 
 polystyrene, 
 with  ${\rm Im}(n_{\rm sphere})= 0.0011$ \cite{Ma03}, 
 scattering dominates over absorption, and the OT has the sign opposite to the input spin AM for most of the range of radii shown in Fig.~\ref{plot_kappa} (solid line).
 We conclude that the scattered field 
 carries AM in excess of  the incident  trapping beam AM  in such cases, thus leading to a
 negative OT.

For small spheres in the  Rayleigh limit,  $a\ll \lambda_0,$ the OT is dominated by extinction rather than by the angular momentum transfer to 
 the scattered field, so that the OT is positive, as shown in Fig.~\ref{plot_kappa}. 
Thus, small microspheres behave as local probes of the incident beam AM, producing  negligible scattering. As a consequence, the OT 
has the same handedness as the laser beam at the objective entrance port. 

  In the opposite 
 range of large spheres, 
 $a\gg \lambda_0,$ 
 the geometrical optics result is recovered as an average over wave-optical interference oscillations~\cite{Mazolli03}.
  Such oscillations are suppressed when absorption is large (dashed line), but are clearly visible 
  in the case of polystyrene (solid line). 
 For transparent spheres, the OT  vanishes in the geometrical optics limit (see Appendix E of  \cite{Mazolli03}), so that 
 the OT oscillates between positive and negative values. 
At the crossover between the Rayleigh and geometrical optics ranges, the OT develops a remarkable negative peak for transparent materials (solid line). 
 
 As found for the transverse trap stiffness 
  $\kappa_{\rho}= - (\partial F_\rho/ \partial \rho) |_{\rho=0},$
such peak is severely modified by optical aberrations  additional to the spherical aberration
already taken into account in MDSA theory.
Indeed, the astigmatism introduced by  small misalignments
in the experimental setup typically 
 degrades the focal region thus leading to a  reduction of  $\kappa_{\rho}$ \cite{Roichman2006,Dutra2012,Dutra2014} in the peak region,
  located at the crossover between the
 Rayleigh and geometrical optics regions.
The effect of astigmatism on $\kappa_\phi(\phi=\phi_{\rm ast})$ ($\phi_{\rm ast}=$ axis angular position)
 is shown 
 by the dotted curve in Fig.~\ref{plot_kappa}. We have employed  MDSA+ theory with
 the  Zernike
  amplitude $A_{\rm ast}=0.25$ 
 obtained by measurements of the reflected focal spot \cite{Dutra2014}, as detailed  in Appendix B.
 Explicit expressions for the optical force within  MDSA+ are given in Appendix C.

Fig.~\ref{plot_kappa} shows that even a small amount of astigmatism not only suppresses the negative
peak found in the ideal MDSA case, but also 
 changes the sign of $\kappa_\phi$ from negative to positive (and hence with the same handedness as the spin AM at the objective entrance port)
 on the lower side of the peak region, thereby extending the 
 positive Rayleigh region to larger values of radius. 
 Thus, a practical implementation of the OT reversal with polystyrene microspheres requires astigmatism to be sufficiently small, and 
microsphere radii in the range $0.4\,\mu{\rm m} \stackrel{<}{\scriptscriptstyle\sim}a \stackrel{<}{\scriptscriptstyle\sim}2\, \mu{\rm  m}.$
We have 
measured the  rotation angle $\alpha$ produced by the 
 OT  for two sphere sizes in this range.
 
\begin{figure} [h]
	\begin{center}
		\includegraphics[width=8.8cm]{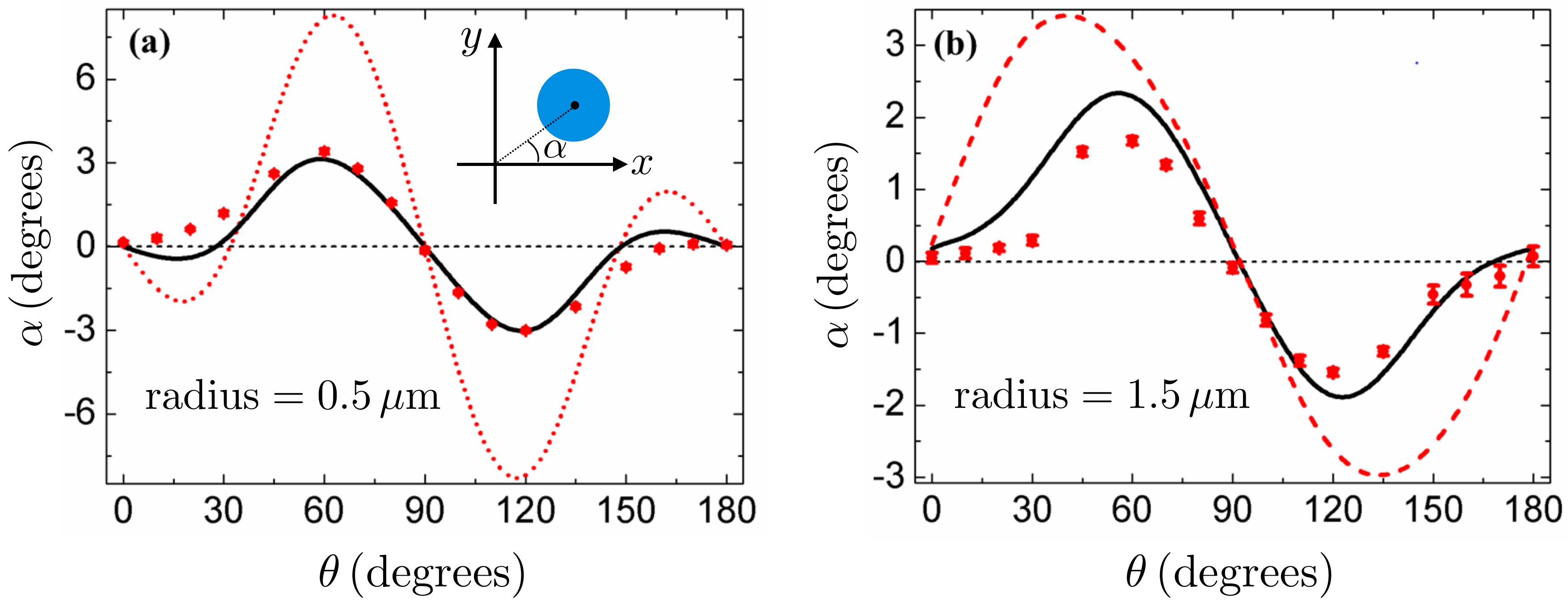}\end{center}
	\caption{Microsphere rotation angle $\alpha$ versus quarter waveplate angle $\theta.$
	Circles with error bars: experimental data. Solid line: MDSA+ theory (no fitting) and 
	microsphere radii (a) $0.5\, \mu {\rm m}$ and (b) $1.5\, \mu {\rm m}.$  	 The dotted line in (a) corresponds to MDSA (zero astigmatism), while the dashed line in (b) is computed with MDSA+ neglecting absorption. 
	}
	\label{plot_alpha}
\end{figure}

In Fig.~\ref{plot_alpha}, we plot 
$\alpha$ versus the QW angle  $\theta.$
Since the off-axis displacements and the rotation angles are  small, we can approximate
\(
\alpha \approx \kappa_\phi/\kappa_\rho.
\) 
When computing $\kappa_\phi$ from Eq.~(\ref{Fphi}), we take $\phi=0$ since the Stokes force is applied along the $x$ direction. 
Within MDSA+, a more involved expression is used (see Supplement 1).
Because of  spherical aberration, the theoretical predictions depend of the focal height from the glass slide. 
We first find the focal height leading to an equilibrium position such that the  microsphere is touching the glass slide.
Then, we add  the height corresponding to the objective upward displacement in order to mimic the experimental procedure \cite{Dutra2014}.
Fig.~\ref{plot_alpha}a  shows experimental results for a microsphere of radius  $a=(0.50\pm 0.02)\,\mu{\rm m},$ together with the theoretical curves based either on 
MDSA (dotted line) or MDSA+ (solid line) with the aforementioned
 astigmatism amplitude and $\phi_{\rm ast}=1^{\rm o}$
  measured for our setup. According to Eq.~(\ref{Fphi}), 
the theoretical curves deviate from a pure sinusoidal function proportional to $\sin 2\theta,$ which is directly associated to the transfer of spin AM to the microsphere, 
because of the cross $\sigma^+\cdot\sigma^-$ term associated to the focal spot asymmetry, which is proportional to $\sin 4\theta.$ The relative contribution of 
the spot asymmetry is smaller when astigmatism is taken into account, as expected since optical aberrations degrade the focal region and thus tend to average out  
the energy density distribution. 

We find  overall good agreement with MDSA+ theory, with no fitting parameters.
The residual disagreement with the experimental data shown in Fig.~\ref{plot_alpha}(a) indicates that our theory slightly overestimates the spot asymmetry effect, so that the experimental curve is closer to a pure sinusoidal function proportional to $\sin 2\theta$ than the MDSA+ one. In particular, the latter predicts a (small) 
rotation with the same handedness as the elliptical polarization for 
$0^{\rm o}<\theta\stackrel{<}{\scriptscriptstyle\sim} 30^{\rm o}$ and $150^{\rm o}\stackrel{<}{\scriptscriptstyle\sim}\theta<180^{\rm o}.$ However,  the experimental data display a rotation opposite to the polarization handedness (given by the sign of $-\sin2\theta$)
for all measured values of $\theta.$

In  Fig.~\ref{plot_alpha}(b), both 
theoretical curves are calculated with
 MDSA+ for the microsphere radius $1.5\, \mu {\rm m}.$
In addition to the solid line corresponding to polystyrene, we also show the result for
 zero absorption (dashed). Although the sphere diameter is much smaller than the attenuation length $\approx 77\,\mu{\rm m}$ for polystyrene, 
 Fig.~\ref{plot_alpha}(b) shows that absorption has a significant  effect on the rotation angle (while it is irrelevant for the  $0.5\, \mu {\rm m}$ microsphere). 
 As expected, absorption tends to decrease the OT reversal. 
 When compared to experimental data, the theoretical prediction based on the 
 value   ${\rm Im}(n_{\rm sphere})=0.0011$ reported by~\cite{Ma03} slightly overestimates the magnitude of 
 $\alpha.$  

\section{Conclusion}

In conclusion,  we have demonstrated a negative OT on trapped
polystyrene microspheres by measuring the rotation of the particle equilibrium position under the effect of a Stokes drag force. 
Good agreement was found between our experimental data and MDSA+ theoretical predictions for the rotation angle, with no fitting. 
 Our method could be applied to characterize the absorption of the trapped sphere material, a property  difficult  to access experimentally, as well as the degree of astigmatism present in the optical setup, given that the microsphere rotation angle is very sensitive to those parameters. 

The negative OT can be seen as an example of  
 spin-orbit interaction  (see~\cite{Bliokh2015} for a recent review).
 Our experiment indeed involves two well-known mechanisms for spin-orbit coupling: 
nonparaxial focusing \cite{Zhao2007,Monteiro2009} and 
  scattering \cite{Bliokh2011}. 
However, there is an important distinction between these two effects. 
 Whereas focusing conserves the total optical AM \cite{Monteiro2009,Bliokh2011}, 
 scattering by the microsphere laterally displaced with respect to the beam symmetry axis 
generates the optical AM excess which is responsible for the negative OT on the sphere center-of-mass.
Since the scattered spin AM cannot be further enhanced with respect to the
spin AM of our circularly-polarized  paraxial Gaussian beam at the the objective entrance port, 
the excess AM is necessarily orbital. 

As a perspective for future work, employing Laguerre-Gauss vortex beams 
 at the objective entrance port might open the way for 
a richer environment, where incident orbital and
spin  AM would be simultaneously set to play. 
Moreover,  employing a laser beam with a larger AM per photon 
might enhance the magnitude of 
the negative OT.

We are grateful to Y. A. Ayala, 
D. S. Ether Jr, P. B. Monteiro and B. Pontes for discussions. 

\textbf{Funding.} National Council for Scientific and Technological Development (CNPq);  Coordination for the Improvement of Higher Education Personnel (CAPES); 
National Institute of Science and Technology Complex Fluids (INCT-FCx);
and Research Foundations of the States of Minas Gerais (FAPEMIG), Rio de Janeiro (FAPERJ) and S\~ao Paulo (FAPESP) (2014/50983-3).

\begin{appendices}

\section{MDSA theory of the optical force in optical tweezers}

We first
define the dimensionless efficiency factor 
 \cite{Ashkin}
\begin{equation}
{\bf Q}(\rho,\phi,z) = \frac{{\bf F}}{n_{\rm w}P/c}
\end{equation}
where ${\bf F}$ is the optical force at position $(\rho,\phi,z)$ in cylindrical coordinates,  $P$ is the laser beam power at the sample region, $n_{\rm w}$ is the refractive index of the immersion fluid (water in our setup) and $c$ is the speed of light. The position of the 
microsphere center (in cylindrical coordinates) $(\rho,\phi,z)$ is measured with respect to the paraxial focus.

There are two separate contributions to the efficiency factor:
\begin{equation}\label{se}
{\bf Q} = {\bf Q}_{\rm s}+{\bf Q}_{\rm e}.
\end{equation}
${\bf Q}_{\rm e}$ is the rate of momentum removal from the incident trapping beam, whereas 
$-{\bf Q}_{\rm s}$ represents the rate of momentum transfer to the scattered field. 

We compute the efficiency factor 
for the elliptical polarization produced by a quarter-wave plate whose fast axis makes an angle $\theta$ with the 
direction of linear polarization of the incident laser beam. 
The azimuthal component of ${\bf Q}$ contains a term proportional to the helicity 
\begin{equation}
\sum_{\sigma=+,-}\, \sigma\, |(1-ie^{-2i\sigma\theta})/2|^2= -\sin(2\theta).
\end{equation}
and a second term associated to the spot asymmetry when the polarization is not circular:
\begin{equation}\label{Qphimain}
Q_{\phi}(\rho,\phi,z) = - Q_{\sigma^+}(\rho,z)\sin2\theta + Q_{\rm cr}(\rho,z)\cos2\theta\sin[2(\phi-\theta)],
\end{equation}
\smallskip
$Q_{\sigma^+}$ corresponds to $\sigma^+$ circularly-polarized trapping beams. Its  partial-wave expansion is given in~\cite{Viana2007}. 

As in the case of linear polarization \cite{Dutra2007}, the 
second term in the r.-h.-s. of (\ref{Qphimain}) results from the
 cross contribution $\sigma^+\cdot \sigma^-$ 
that appears when the polarization is not circular. 
Following 
(\ref{se}), 
we write  $ Q_{\rm cr} = Q^{\rm(cr)}_{{\rm s}}+Q^{\rm(cr)}_{{\rm e}}.$
The scattering contribution is given by
\begin{widetext}
\[
Q^{\rm(cr)}_{{\rm s}}(\rho,z) = -\frac{4\gamma^2}{AN}\sum_{j=1}^{\infty}\sum_{m=-j}^j
\frac{\sqrt{j+m+1}}{j+1}
\,{\rm Im}\Biggl[\sqrt{j(j+2)(j+m+2)}(a_{j}a_{j+1}^{*}-b_{j}b_{j+1}^{*})\times\]
\begin{equation}\label{Qcr}
(G_{j,-m}G_{j+1,m+1}^{*} - G_{j,m}G_{j+1,-m-1}^{*})
-\,\frac{(2j+1)\sqrt{(j-m)}}{j}\,a_{j}b_{j}^{*}\,
Re(G_{j,m}G_{j,-(m+1)}^{*}+G_{j,m+1}G_{j,-m}^{*})\Biggl].
\end{equation}
The extinction term is written as 
\begin{equation}\label{Qcre}
Q^{\rm (cr)}_{\rm e}(\rho,z)=-\frac{2\gamma^2}{AN}\,{\rm Im}\sum_{j=1}^{\infty}\sum_{m=-j}^j(2j+1)(a_{j}-b_{j})
G_{j,m}\left(G_{j,-(m+1)}^{+}+G_{j,-(m-1)}^{-}\right)^*.
\end{equation}
The fraction of  power transmitted into the sample chamber is given by
\begin{equation}
{A} = 16\gamma^2   \label{A}
\int_0^{\sin\theta_0} ds\, s \,\exp(-2\gamma^2s^2)\, \frac{\sqrt{\left( 1 - s^2 \right)\left(N^2- 
s^2\right)}}{\left(\sqrt{1- s^2} + \sqrt{ N^2- s^2}\right)^2}
\end{equation} 
with $\sin\theta_0=\min\{N,\mbox{NA}/n_{\rm glass}\}$ and $N=n_{\rm w}/n_{\rm glass}$ ($n_{\rm glass}=$ refractive index of the glass slide). 
The parameter $\gamma = f/w$ is the ratio between the objective focal length 
and the beam waist at the entrance aperture of the objective. The Mie coefficients $a_j$ and $b_j$ represent the scattering amplitudes for electric and magnetic multipoles, respectively
\cite{Bohren&Huffman}.

The multipole coefficients appearing in (\ref{Qcr})  are given by: 
\begin{equation}
G_{j,m}=\int_{0}^{\theta_{0}}d\theta_k\sin{\theta_k}\sqrt{\cos\theta_k}e^{-\gamma^2\sin^2\theta_k}
d_{m,1}^{j}(\theta_{\rm w})T(\theta)J_{m-1}(k \rho\sin\theta_k)e^{i\Phi_{\rm g-w}(\theta_k)}e^{ik_{\rm w}z\cos\theta_{\rm w}},
\end{equation}
\end{widetext}
with 
$k=2\pi n_{\rm glass}/\lambda_0,$ $k_{\rm w}=Nk,$ and 
$\sin\theta_{\rm w}= \sin{\theta_k}/N.$ $d_{m,1}^{j}(\theta_{\rm w})$ are the Wigner finite rotation matrix elements \cite{Edmonds}
 evaluated at the 
angle $\theta_{\rm w}$ in the immersion fluid and $J_{m}(x)$ are the cylindrical Bessel functions \cite{Watson}. 
Refraction at the interface between the glass slide and the immersion fluid  introduces the 
 Fresnel transmission amplitude 
\begin{equation}
T(\theta_k)= \frac{2\cos\theta_k}{\cos\theta_k+N\cos\theta_{\rm w}}
\end{equation}
and the
 spherical aberration phase \cite{Torok95} ($L=$ height of paraxial focus with respect to the glass slide)
\begin{equation}
\Phi_{\rm g-w}(\theta_k)=k\left( -\frac{L}{N}\cos\theta_k+NL\cos\theta_{\rm w}\right). \label{aberration_interface}
\end{equation}

The extinction contribution (\ref{Qcre}) 
 is written in terms of the coefficients

\begin{eqnarray}
G_{j,m}^{\pm}&=&\int_{0}^{\theta_{0}}d\theta\sin\theta_k\sin\theta_{\rm w}\sqrt{\cos \theta_k}e^{-\gamma^2\sin^2\theta_k}
d_{m\pm1,1}^{j}(\theta_{\rm w})\times \nonumber\\
&& T(\theta_k)J_{m-1}(k\rho\sin\theta_k)e^{i\Phi_{\rm g-w}(\theta_k)}e^{ik_{\rm w}z\cos\theta_{\rm w}}.
\end{eqnarray}

\section{Characterization of Primary Astigmatism }

Small misalignments in our optical system introduce astigmatism into the trapping beam.
As a consequence, the focused beam angular spectrum contains a Zernike phase \cite{Born&Wolf} 
\begin{equation}\label{s1}
\Phi_{\rm ast}(\theta_k,\phi_k) = 2\pi A_{\rm ast}\biggl(\frac{\sin\theta_k}{\sin\theta_{0}}\biggr)^2\cos [2(\phi_k-\phi_{\rm ast})].
\end{equation} 
We characterize  the amplitude $A_{\rm ast}$ and the axis angular position $\phi_{\rm ast}$  in our setup by employing the  method described in  \cite{Dutra2014}. We replace our sample by a mirror 
that  reflects the laser beam back towards  the objective. 
The laser spot image
 is conjugated  by the microscope tube lens 
 onto a
 a CMOS camera (Hamamatsu Orca-Flash2.8 C11440-10C) for data analysis.
  We employ the piezoelectric nanopositioning system PI (Digital Piezo Controller E-710, Physik Instrumente, Germany)  to move the
  mirror  across the focal region with controlled velocity $V=100\,{\rm nm/s}.$ 

\begin{figure} [htbp]
	\centering
	\includegraphics[scale=0.3]{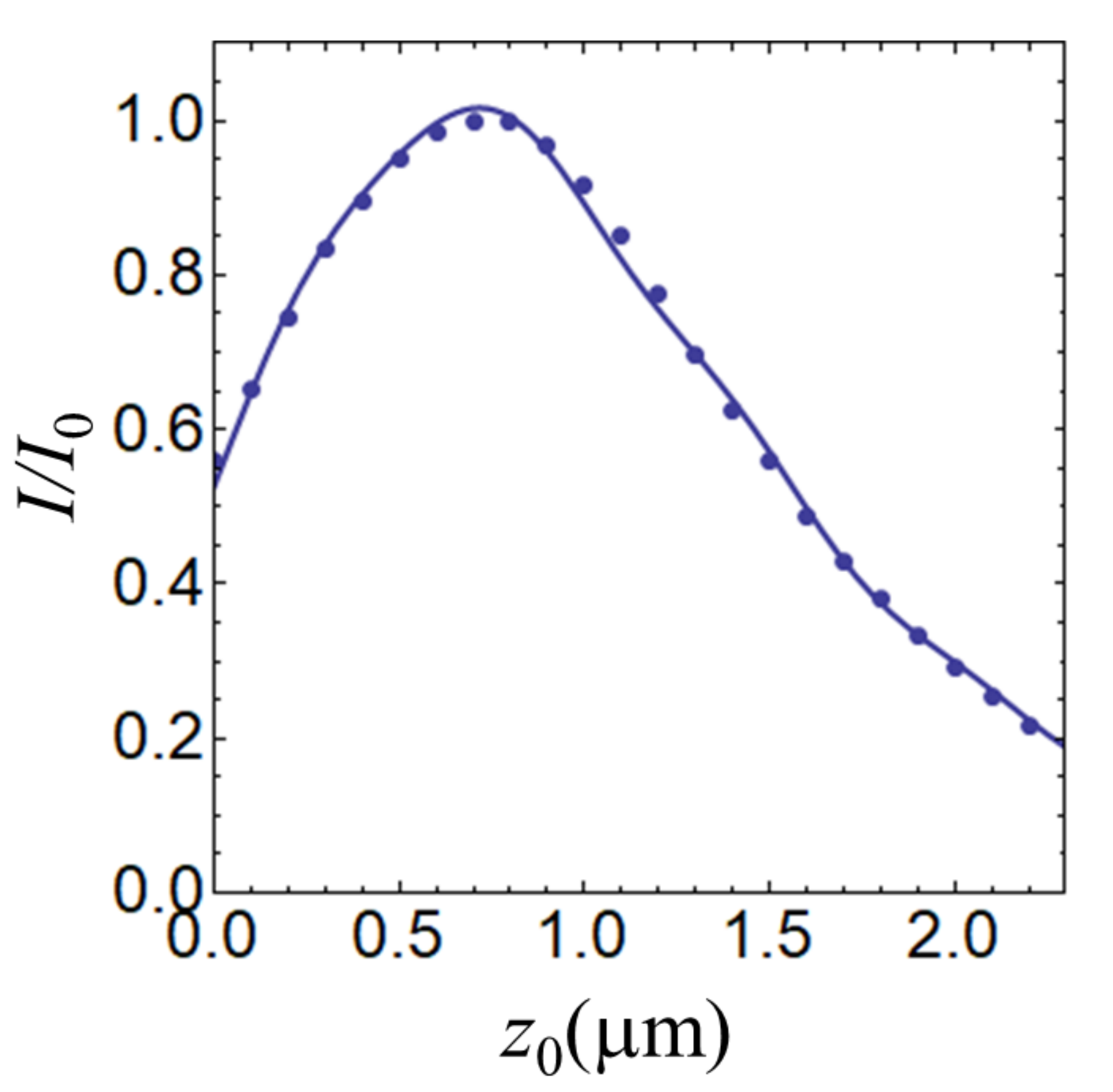}
	\caption{Normalized intensity at the spot center as a function of the mirror axial position $z_{0}:$ 
	 experimental data (points) and  theoretical model (solid line). The astigmatism amplitude $A_{\rm ast} = 0.25 \pm 0.04$ is found as a fitting parameter. }
	\label{plot_astigm} 
\end{figure}

The spot changes as a function of the mirror position $z_0.$ $\phi_{\rm ast}$ is 
defined as the angle between the  spot major axis and the $x$ axis when the mirror is placed at the tangential focus. We also measure the value of 
$\phi_{\rm ast}+90^{\rm o}$ by considering the direction of the major axis when the  mirror is placed at the sagittal focus. 
We have analyzed five  images at each focus and found $\phi_{\rm ast}= (1 \pm 1)^{\rm o}.$ 

To determine the astigmatism amplitude $A_{\rm ast},$ we fit (least square method) the 
 light intensity
 at the center of the spot as function of  $z_0,$
 as shown in Fig.~\ref{plot_astigm}.
  The theoretical curve is 
 obtained from the nonparaxial propagation through the optical system~\cite{Novotny01} as described in detail elsewhere~\cite{Dutra2014}.
 We find  $A_{\rm ast} = 0.25 \pm 0.04.$

\section{ MDSA+ theory }

Once the astigmatism parameters are characterized (see previous section), we are able to compute the optical force by generalizing MDSA theory in order to take 
such primary aberration into account (MDSA+) \cite{Dutra2014}. Since the astigmatism phase 
(\ref{s1}) is not rotationally symmetric around the axis, 
Eq.~(\ref{Qphimain}) no longer holds, and 
the optical force field has a more complex dependence on the microsphere angular  coordinate $\phi.$

Explicit expressions for the cylindrical components are given below.
The microsphere position $(\rho,\phi,z)$ is now measured  with respect to the diffraction focus. 
 We employ the notation
\[
\sum_{jm\sigma} \equiv \sum_{j=1}^{\infty}\sum_{m=-j}^j \sum_{\sigma=-1,1}
\]

\begin{widetext}
\textbf{Scattering force}

\[
Q_{s z}= -\frac{4\gamma^2}{AN}{\rm Re}\sum_{jm\sigma}\frac{\sqrt{j(j+2)(j+m+1)(j-m+1)}}{j+1}\times \nonumber\\
  \biggl[(a_{j}a_{j+1}^{*}+b_{j}b_{j+1}^{*})\times\]\[
G^{(\sigma)}_{j,m}G^{(\sigma)*}_{j+1,m}(1-\sigma\sin2\theta)+(a_{j}a_{j+1}^{*}-b_{j}b_{j+1}^{*})G^{(\sigma)}_{j,m}G^{(-\sigma)*}_{j+1,m}\cos2\theta \,e^{i2\sigma(\phi-\theta)}
\biggr] 
\]\begin{eqnarray}
 -\frac{4\gamma^2}{AN}{\rm Re}\sum_{jm\sigma}
\frac{(2j+1)}{j(j+1)}m\sigma a_{j}b_{j}^{*}\left(\vert G^{(\sigma)}_{j,m}\vert^2(1-\sigma\sin2\theta)-G^{(\sigma)}_{j,m}G^{(-\sigma)}_{j,m}{}^*\cos2\theta \,e^{i2\sigma(\phi-\theta)}\right), \label{Qszp}
\end{eqnarray}

\[
Q_{s\rho} = \frac{2\gamma^2}{AN}\sum_{jm\sigma}\frac{\sqrt{j(j+2)(j+m+1)(j+m+2)}}{j+1}\times\\
 {\rm Im}\biggl\lbrace
(a_{j}a_{j+1}^{*}+b_{j}b_{j+1}^{*})\times\]\[
\left[
G^{(\sigma)}_{j,m}G^{(\sigma)*}_{j+1,m+1}+G^{(\sigma)}_{j,-m}G^{(\sigma)*}_{j+1,-(m+1)}\right](1-\sigma\sin2\theta)\\
+(a_{j}a_{j+1}^{*}-b_{j}b_{j+1}^{*})  
\left[G^{(\sigma)}_{j,m}G^{(-\sigma)*}_{j+1,m+1}+G^{(\sigma)}_{j,-m}G^{(-\sigma)*}_{j+1,-(m+1)}\right]\cos2\theta\, e^{i2\sigma(\phi-\theta)}
\biggl\rbrace  \] \[
-\frac{4\gamma^2}{AN}\sum_{jm\sigma}
 \frac{(2j+1)}{j(j+1)}\sigma\sqrt{(j-m)(j+m+1)}\biggl[{\rm Re}(a_{j}b_{j}^{*})
{\rm Im}( G^{(\sigma)}_{j,m}G^{(\sigma)*}_{j,m+1}(1-\sigma\sin2\theta))+\]\begin{equation} 
 {\rm Im}(a_{j}b_{j}^{*}){\rm Re}(G^{(\sigma)}_{j,m+1}G^{(-\sigma)*}_{j,m}\cos2\theta\, e^{i2\sigma(\phi-\theta)})\biggr]       \label{Qsrho}
\end{equation}

\[
Q_{s\phi} = -\frac{2\gamma^2}{AN}\sum_{jm\sigma}\frac{\sqrt{j(j+2)(j+m+1)(j+m+2)}}{j+1}\times\\
 {\rm Re}\biggl\lbrace
(a_{j}a_{j+1}^{*}+b_{j}b_{j+1}^{*})\times\]\[
\left[
G^{(\sigma)}_{j,m}G^{(\sigma)*}_{j+1,m+1}-G^{(\sigma)}_{j,-m}G^{(\sigma)*}_{j+1,-(m+1)}\right](1-\sigma\sin2\theta)\\
+(a_{j}a_{j+1}^{*}-b_{j}b_{j+1}^{*})\]\[
\left[G^{(\sigma)}_{j,m}G^{(-\sigma)*}_{j+1,m+1}-G^{(\sigma)}_{j,-m}G^{(-\sigma)*}_{j+1,-(m+1)}\right]\cos2\theta\, e^{i2\sigma(\phi-\theta)}
\biggl\rbrace  \] \[
+\frac{4\gamma^2}{AN}\sum_{jm\sigma}
 \frac{(2j+1)}{j(j+1)}\sigma\sqrt{(j-m)(j+m+1)}\biggl[{\rm Re}(a_{j}b_{j}^{*})
{\rm Re}( G^{(\sigma)}_{j,m}G^{(\sigma)*}_{j,m+1}(1-\sigma\sin2\theta))+\]\begin{equation} 
 {\rm Im}(a_{j}b_{j}^{*}){\rm Im}(G^{(\sigma)}_{j,m+1}G^{(-\sigma)*}_{j,m}\cos2\theta\, e^{i2\sigma(\phi-\theta)})\biggr].       \label{Qsphi}
\end{equation} 

 \textbf{Extinction force}

\begin{equation}
Q_{ez}=\frac{2\gamma^2}{AN}{\rm Re}\sum_{jm\sigma}(2j+1)
G^{(\sigma)}_{j,m}
\left[(a_{j}+b_{j})
G_{j,m}^{C,(\sigma)}{}^*(1-\sigma\sin2\theta)
+(a_{j}-b_{j})G_{j,m}^{C,(-\sigma)}{}^* \cos2\theta\, e^{i2\sigma(\phi-\theta)}\right],
\label{Qez}
\end{equation}

\[
Q_{e\rho}=\frac{\gamma^2} {AN}{\rm Im}\sum_{jm\sigma}(2j+1)G^{(\sigma)}_{j,m}
\biggl[(a_{j}+b_{j}) 
\left(G^{-,(\sigma)}_{j,m+1} - G^{+,(\sigma)}_{j,m-1}\right)^*(1-\sigma\sin2\theta)\]\begin{equation}
+(a_{j}-b_{j})
\left(G^{-,(-\sigma)}_{j,m+1} - G^{+,(-\sigma)}_{j,m-1}\right)^*\cos2\theta\, e^{i2\sigma(\phi-\theta)} \biggr]
 \label{Qerho}
\end{equation}

\[
Q_{e\phi}=-\frac{\gamma^2} {AN}{\rm Re}\sum_{jm\sigma}(2j+1)G^{(\sigma)}_{j,m}
\biggl[(a_{j}+b_{j}) 
\left(G^{+,(\sigma)}_{j,m-1}+G^{-,(\sigma)}_{j,m+1}\right)^*(1-\sigma\sin2\theta)\]\begin{equation}
+(a_{j}-b_{j})
\left(G^{+,(-\sigma)}_{j,m-1}+G^{-,(-\sigma)}_{j,m+1}\right)^*\cos2\theta\, e^{i2\sigma(\phi-\theta)} \biggr],
 \label{Qerho}
\end{equation}

\textbf{ Multipole coefficients}

\begin{equation}\label{Gjm}
G^{(\sigma)}_{jm}(\rho,\phi,z)=\int_{0}^{\theta_0}d\theta_{k}\sin\theta_{k}\sqrt{\cos \theta_{k}}\,e^{-\gamma^2\sin^2\theta_{k}}\,
T(\theta_{k})\,d_{m,\sigma}^{j}(\theta_{\rm w})\,f^{(\sigma)}_{m}(\rho,\phi,\theta_{k})\,
e^{i[\Phi_{\rm g-w}(\theta_{k})+k_{\rm w}\cos\theta_{\rm w} z ]},
\end{equation}

\begin{equation}
G_{j,m}^{C,(\sigma)}(\rho,\phi,z)=\int_{0}^{\theta_0}d\theta_{k}\sin\theta_{k} \cos\theta_{\rm w} \sqrt{\cos \theta_{k}}\,e^{-\gamma^2\sin^2\theta_{k}}T(\theta_{k})d_{m,\sigma}^{j}(\theta_{\rm w})f^{(\sigma)}_{m}(\rho,\phi,\theta_{k})e^{i[\Phi_{\rm g-w}(\theta)+k_{\rm w}\cos\theta_{\rm w} z ]} \label{multipole coefficient}
\end{equation}

\begin{equation}
G_{j,m}^{\pm,(\sigma)}(\rho,\phi,z)= \int_{0}^{\theta_0}d\theta_{k}\sin\theta_{k}\sin\theta_{\rm w}
\sqrt{\cos \theta_{k}}\,e^{-\gamma^2\sin^2\theta_{k}}T(\theta_{k})d_{m\pm1,\sigma}^{j}(\theta_{\rm w})f^{(\sigma)}_{m}(\rho,\phi,\theta_{k}) e^{i[\Phi_{\rm g-w}(\theta)+k_{\rm w}\cos\theta_{\rm w} z ]},
\end{equation}

For astigmatic beams, the multipole coefficients written above contain an additional series involving cylindrical Bessel functions:
\begin{equation}
f^{(\sigma)}_{m}(\rho,\phi,\theta_{k}) =\sum_{s=-\infty}^{\infty}(-i)^s 
J_{s}\biggl(2\pi A_{\rm ast} \frac{\sin^2\theta_{k}}{\sin^2\theta_{0}}\biggl)\,
J_{2s+m-\sigma}(k\rho\sin\theta_{k})\,
e^{2is(\phi_{\rm ast}-\phi)}.
 \label{function_f}
\end{equation}
\end{widetext}

\end{appendices}

%%%%%%%%%%%%%%%%%%%%%%% References %%%%%%%%%%%%%%%%%%%%%%%%%

\bibliography{biblio_spin}

\end{document}